\documentclass[a4paper,fleqn,usenatbib]{mnras}
\usepackage{amssymb,amsmath,graphicx,psfrag,mathtools}
\usepackage[T1]{fontenc} 
\usepackage{ae,aecompl} 
\usepackage{url}
\usepackage{revsymb}
\hypersetup{draft} 
\title[Periodically Modulated FRB]{Periodically Modulated FRB as Extreme
Mass Ratio Binaries}
\author[J. I. Katz]{
J. I. Katz,$^{1}$\thanks{E-mail katz@wuphys.wustl.edu} 
\\
$^{1}$Department of Physics and McDonnell Center for the Space Sciences,
Washington University, St. Louis, Mo. 63130 USA 
}
\date{Accepted XXX.  Received YYY; in original form ZZZ} 
\pubyear{2023} 
\date{\today}
\begin{document} 
\label{firstpage} 
\pagerange{\pageref{firstpage}--\pageref{lastpage}} 
\maketitle 
\begin{abstract}
	The activity of at least one repeating Fast Radio Burst (FRB) source
	is periodically modulated.  If this modulation is the result of
	precession of the rotation axis and throat of an accretion disc
	around a black hole, driven by a companion that is also the source
	of accreted mass, then it may be possible to constrain the mass of
	the black hole.  The dynamics is analogous to that of superorbital
	periods in ordinary mass-transfer binaries in which the accreting
	object may be a stellar-mass black hole, a neutron star or a white
	dwarf, but in the FRB source it may be an intermediate mass black
	hole.  In a semi-detached (mass-transferring) binary the orbital
	period is related to the mean density of the mass-losing star.
	Assuming a value for its density and identifying the observed
	modulation period as a disc precession period would determine the
	mass ratio and the mass of the black hole.  This model and
	magnetar-SNR models make distinguishable predictions of the
	evolution of the FRB rotation measure that may soon be tested in
	FRB 121102. 
\end{abstract}
\section{Introduction}
Fast radio burst sources may be divided into two classes, frequent
repeaters, with intervals observed by the most sensitive telescopes of
$\sim 1\,$min during periods of intense activity, and apparent
non-repeaters.  These classes differ in other properties \citep{K22a} and
may be disjoint.  The apparent non-repeaters likely repeat at long intervals
because the rate of catastrophic events appears to be insufficient to
explain them \citep{H20}.  \citet{K17,S21} suggested that the frequent
repeaters are produced by accretion discs around intermediate mass
($10^2$--$10^6\,M_\odot$) black holes.  This hypothesis is supported by the
explanation of the periodic modulation, and its jitter around strict
periodicity, of the activity of FRB 2018916B as a result of disc precession
\citep{K22b}.

In such a model the binary system is semi-detached, or nearly so; the
nondegenerate companion is in (or very close to) contact with its Roche lobe
and mass transfer is continuous, although its rate may vary.  The bursts
themselves must be the result of brief plasma storms superposed on the more
slowly varying disc orientation and mass flow rate.

Precession of the disc angular momentum axis about the orbital axis is
the cause of many or most of the ``superorbital'' periods ubiquitous in
mass-transfer binaries.  Precession is a weakly damped eigenmode of
dynamical excitation about the disc's lowest energy state, in which it is
circular and coplanar with the orbit.  Excitation of this mode is generally
attributed to ``turbulence'' (fluctuating torques) in the accretion disc or
fluctuating deviations of the mass-transfer stream from the orbital plane,
perhaps driven by turbulence or spin mis-alignment in the mass-losing star.
The amplitude of excitation cannot be predicted, but in the most favorable
systems may be measured, at least approximately.  Precession may be
manifested as periods in which the accreting object is eclipsed by the disc
(Her X-1; \citet{K73,LG82}) or directly measured in Doppler shifts of a
disc's axial jet (SS 433; \citet{K80,M81,K82,M84,F04}).  Four such systems
are well established \citep{L98} with ratios of their precession to orbital
periods in the range 12.5--22:1.  These have primary to secondary mass
ratios ${\cal O}(1)$, though only approximately determined.
\section{Precession Rate}
The precession rate $\omega_{pre}$ of a circular ring of radius $R$ in orbit
around an object of mass $M_p$, that itself has a secondary object of mass
$M_s$ in a circular orbit of radius $a \gg R$ and angular rate
$\Omega_{orb}$, is \citep{K73,L98}
\begin{equation}
	\label{basic}
	{\omega_{pre} \over \Omega_{orb}} = - {3 \over 4} q \left({R \over
	a}\right)^3 \cos{\delta},
\end{equation}
where $q = M_s/M_p$ and $\delta$ is the inclination between the orbital
and ring angular momentum axes ($\cos{\delta}$ is generally taken as unity).
This is a classical result whose origin is Newton's theory of the precession
of the plane (recession of the nodes) of the Moon's orbit under the
gravitational influence of the Sun.

\citet{L98}, making plausible assumptions for the distribution of mass in a
disc extending inward from $R$ to the compact object, calculated that the
factor of $3/4$ in Eq.~\ref{basic} should be replaced by $3/7$ because the
effective value of $R$ for a continuous accretion disc is less than its
outer radius\footnote{{\it N.B.:\/} \citet{L98} used $\mu$ for our $q$.}.
Accretion discs may be accompanied by and coupled to excretion discs
extending beyond the RCR, increasing the coefficient in Eq.~\ref{basic},
offsetting the reduction calculated by \citet{L98}.  At some radius $> R$
but $< a$ the disc is truncated as its orbits become unstable \citep{B74},
limiting the possible value of the coefficient.

$R$ is usually taken as the Roche Circularization Radius (RCR) \citep{K73}
at which a stream of matter spilling over the L1 point from the mass-losing
secondary circularizes its motion by dissipatively colliding with itself,
conserving angular momentum.  In the absence of angular momentum transport
and further viscous dissipation, a circular ring (with $\delta = 0$) is the
lowest energy state.  Viscous dissipation more slowly turns the ring into a
disc.
\section{Extreme Mass Ratios}
Eq.~\ref{basic} is only valid in the limit $R/a \to 0$.  This is a good
approximation if $q\not\ll 1$.  Calculations of binary accretion discs
\citep{WP72,K73,FKR} have usually considered $M_s \gtrsim M_p$\footnote{In
some papers $q \equiv M_p/M_s$, the reciprocal of the definition here.}.
This is appropriate for many binary stars undergoing mass transfer, but not
for extreme recycled binary pulsars and cataclysimic variables in which the
mass-losing star may have a mass of hundredths of $M_\odot$.

A widely used fit to the RCR (Eq.~4.20 of \citet{FKR}) breaks down for
$q < 0.1$, for which it would imply a value greater than the separation
between the stars.  Yet we are concerned with accretion from ordinary stars
onto intermediate mass black holes, so that $q \ll 1$, perhaps by orders of
magnitude.

It is straightforward to calculate the Roche Circularization Radius by
integrating the orbits of freely falling particles (or gas streams with low
sound speeds) from the inner Lagrange Point L1, making the usual Roche
assumption of an aligned and synchronously rotating mass-losing (secondary)
star with a cool (small scale height) atmosphere.  The circularization
criterion 
\begin{equation}
	\label{RCR}
	R = {L_1^2 \over G M_p},
\end{equation}
where $L_1$ is the specific angular momentum about the primary (black hole),
must be evaluated in an inertial frame, although the orbit calculation is
conveniently done in the co-rotating frame.  For $q = {\cal O}(1)$ this is a
small correction, but for $q \ll 1$ it is essential because the circularized
matter orbits the primary at an angular rate that approaches $\Omega_{orb}$
in the limit $q \to 0$.

\begin{figure}
	\centering
	\includegraphics[width=3.3in]{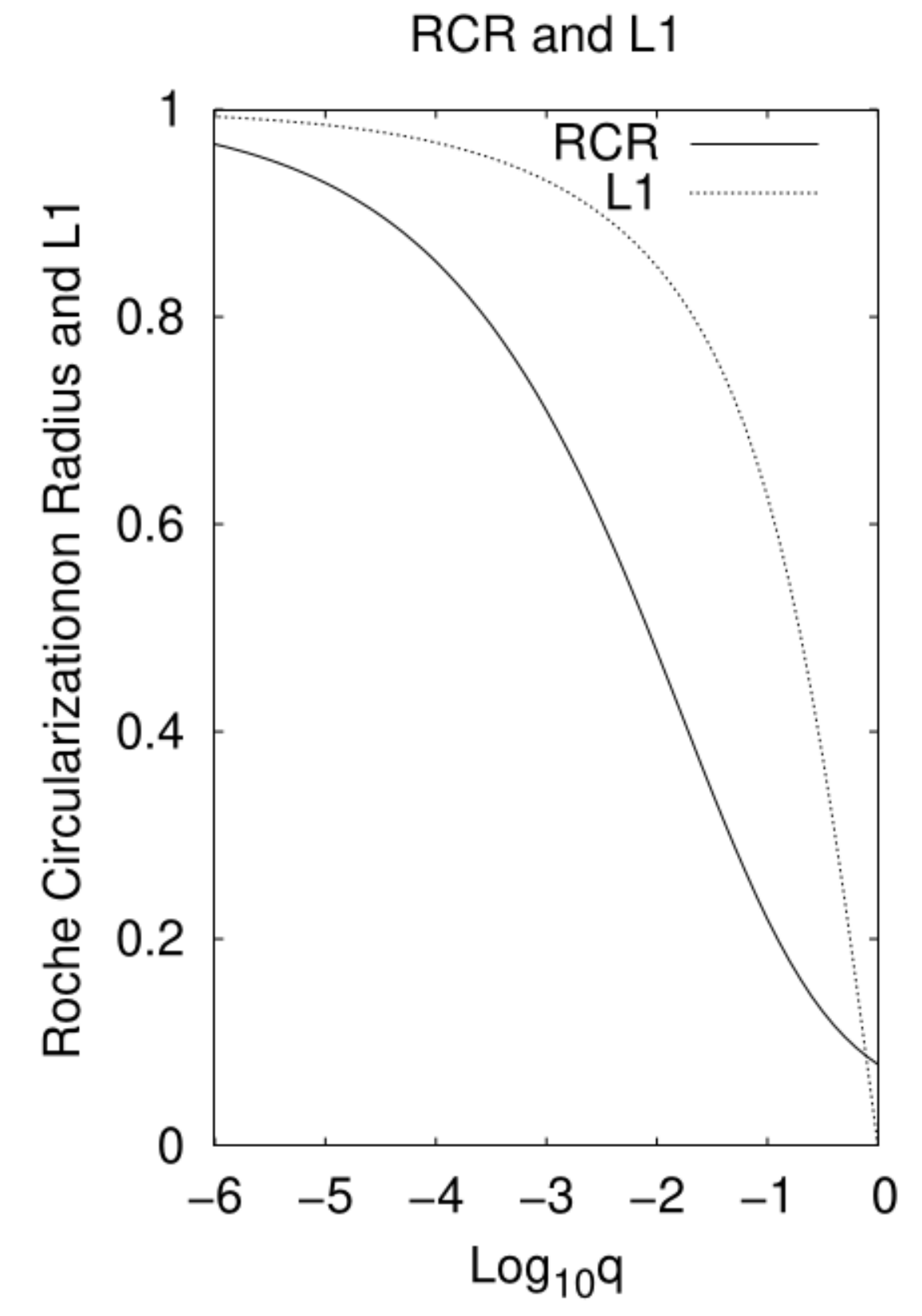}
	\caption{\label{L1} Roche Circularization Radius (RCR) and the
	distance of the inner Lagrange point L1 from the center of mass, in
	units of the binary separation $a$, as functions of the mass ratio
	$q$ for $q \le 1$.}
\end{figure}

\begin{figure}
	\centering
	\includegraphics[width=3.3in]{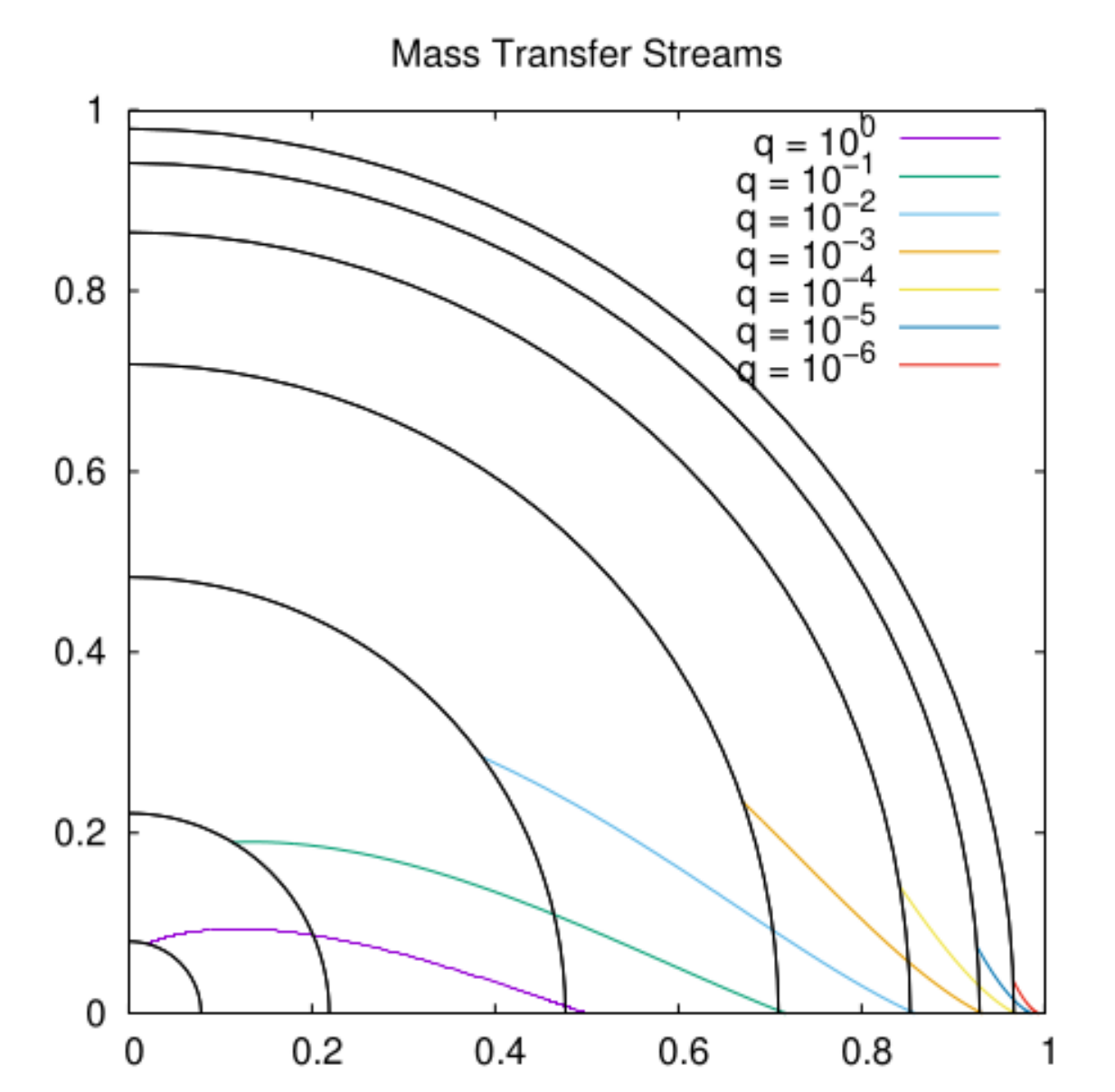}
	\caption{\label{orbits} Orbits of particles or cool matter spilling
	over the secondary's Roche lobe at L1 for several values of $q \ll
	1$.  The paths are truncated at the RCR, and these radii are
	illustrated by circles centered at the primary.  The origin is taken
	at the primary (not the center of mass), and the stellar separation
	is taken as unity.  The paths begin at L1 on the abscissa, and are
	bent by the Coriolis force so it is only necessary to show one
	quadrant.}
\end{figure}

The location of the inner Lagrange point L1 and RCR are shown in
Fig.~\ref{L1} as a function of $q$.  The location of L1, obtained by
numerically finding the maximum of the effective potential (including the
centrifugal force) in the corotating frame, agrees in the limit $q \to 0$
with the asymptotic expression of \citet{K77}.  The orbits of freely falling
particles or cool gas streams from L1 to their RCR are shown in
Fig.~\ref{orbits}.

\begin{figure}
	\centering
	\includegraphics[width=3.3in]{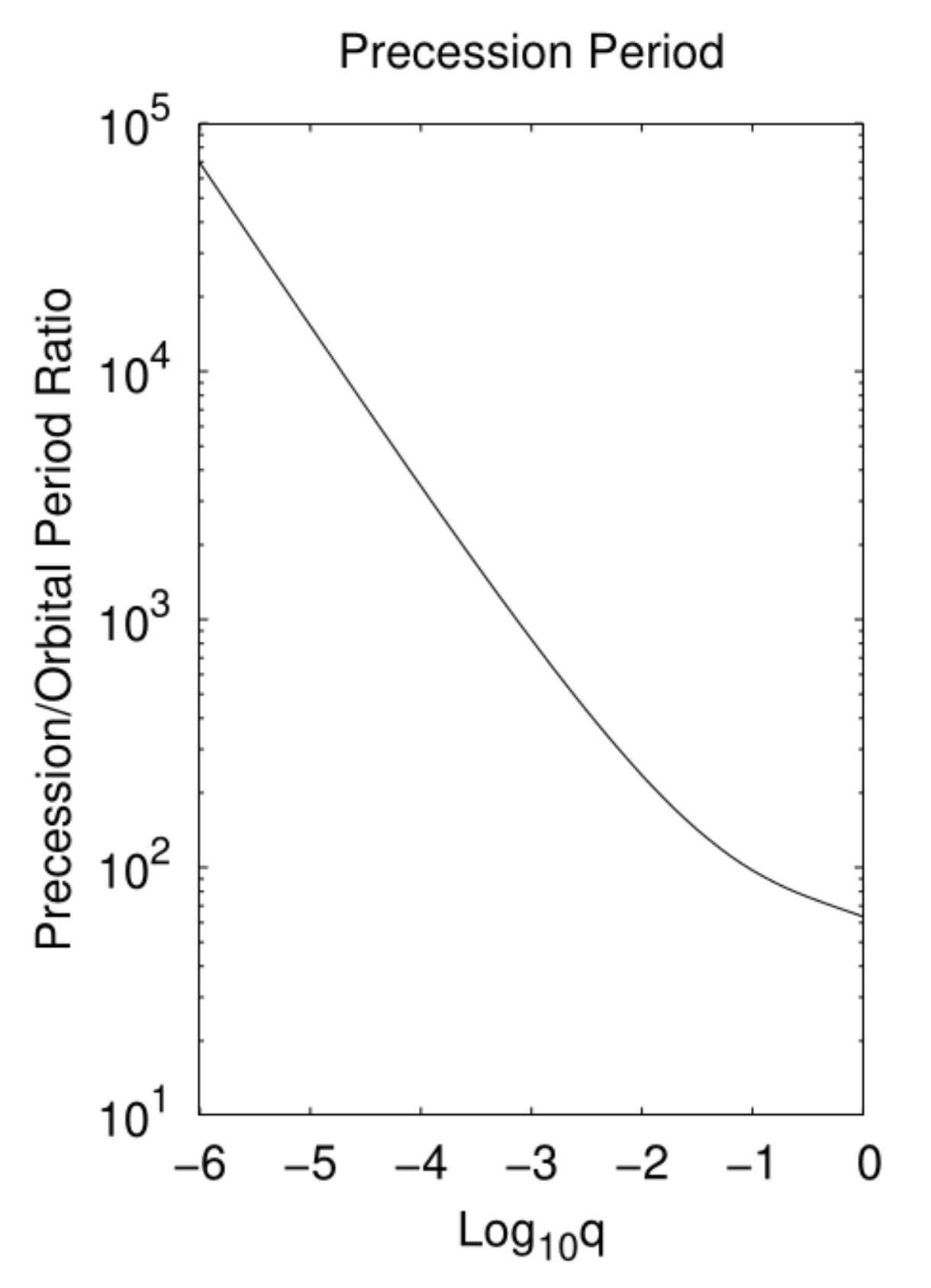}
	\caption{\label{period} The precession period in the Roche model for
	$q \le 1$, as a multiple of the binary orbital period.}
\end{figure}

For $q \ll 1$, $R \not\ll a$ and Eq.~\ref{basic} is not a valid
approximation for the precession rate. Then the precessional torque on an
inclined ring ($\delta \ne 0$) is nearly impulsive, concentrated on mass
elements as they pass near L1 at distances $\ll a$.  For $q \ll 1$ it must
be evaluated by numerical integration.  It is assumed that $\delta \ll
1 - R/a \ll 1$.  The results are shown in Fig.~\ref{period} as the ratio of
the precession period to the binary orbital period in the limit $\delta \to
0$.  In the regime $q \ll 1$ the results are well fit by the expression
\begin{equation}
	\label{periodfit}
	{P_{pre} \over P_{orb}} \approx 6.86 q^{-2/3}.
\end{equation}

\section{Application to Intermediate Mass Black Hole Binaries}
The difficulty of applying Eq.~\ref{periodfit} to observations is that while
it can be argued that in at least one repeating FRB \citep{K22b} $P_{pre}$
is observed as an activity modulation period, neither the value of $P_{orb}$
nor the secondary (mass-losing star) mass $M_s$ is known.  Hence it is not
possible to use Eq.~\ref{periodfit} directly to determine $q$ and the black
hole mass.

We suggest some possible ranges of these values to indicate how the method
may be used if more data become available.  In a fair approximation to Roche
geometry for $q \ll 1$ and $a-R \ll a$
\begin{equation}
	\label{lobes}
	{M_p \over a^3} \approx {M_p \over R^3} \approx {M_s \over
	(a-R)^3} \approx {M_s \over R_s^3},
\end{equation}
where $R_s$ would be the radius of the secondary star were it not perturbed
by the gravitational field of the primary.

Using Kepler's law for a circular orbit, Eq.~\ref{lobes} implies, nearly
independent of $q$,
\begin{equation}
	\label{rho}
	P_{orb} \approx \sqrt{3 \pi \over G \langle \rho \rangle},
\end{equation}
where $\langle \rho \rangle$ is the mean density of the secondary star.

If the secondary is a white dwarf then the orbital period is ${\cal O}
(\text{10 s})$.  This may be excluded because its lifetime to gravitational
radiation would be shorter than the observed multi-year lower bounds on the
lifetimes of repeating FRB sources.

The mean densities of nondegenerate stars range from ${\cal O}
(\text{100 g/cm}^3)$ for low mass main sequence stars to ${\cal O}
(10^{-6}\,\text{g/cm}^3)$ for supergiants.  This is consistent with a very
wide range of $P_{orb}$ and an even wider range of $P_{pre}$ (because $q$ is
not known {\it a priori\/}).  Very long $P_{pre}$ are unobservable.
\section{Secondary Density and Black Hole Masses}
If the 16.3 day periodic modulation of the activity of FRB 180916B
\citep{P21} is interpreted \citep{K22b} as a disc
precession period, a coupled constraint on $q$ and $\langle \rho \rangle$
can be set.  Combining Eqs.~\ref{periodfit} and \ref{rho},
\begin{equation}
	\label{constraint}
	\langle \rho \rangle \approx 0.3 q^{-4/3}\ \text{g-cm}^{-3}.
\end{equation}
For $q \lesssim 0.2$ (a necessary condition for the applicability of
Eq.~\ref{constraint}), $\langle \rho \rangle \gtrsim 3$ g-cm$^{-3}$; the
secondary star must be a low mass main-sequence star; see Eq.~\ref{density}.
The same interpretation of the 159 day periodicity of FRB 121102 \citep{R20}
implies $\langle \rho \rangle \gtrsim 3.5 \times 10^{-5} q^{-4/3}$
g-cm$^{-3}$, consistent with $q \ll 0.2$, a subgiant secondary star, or
both.

\begin{figure}
	\centering
	\includegraphics[width=3.3in]{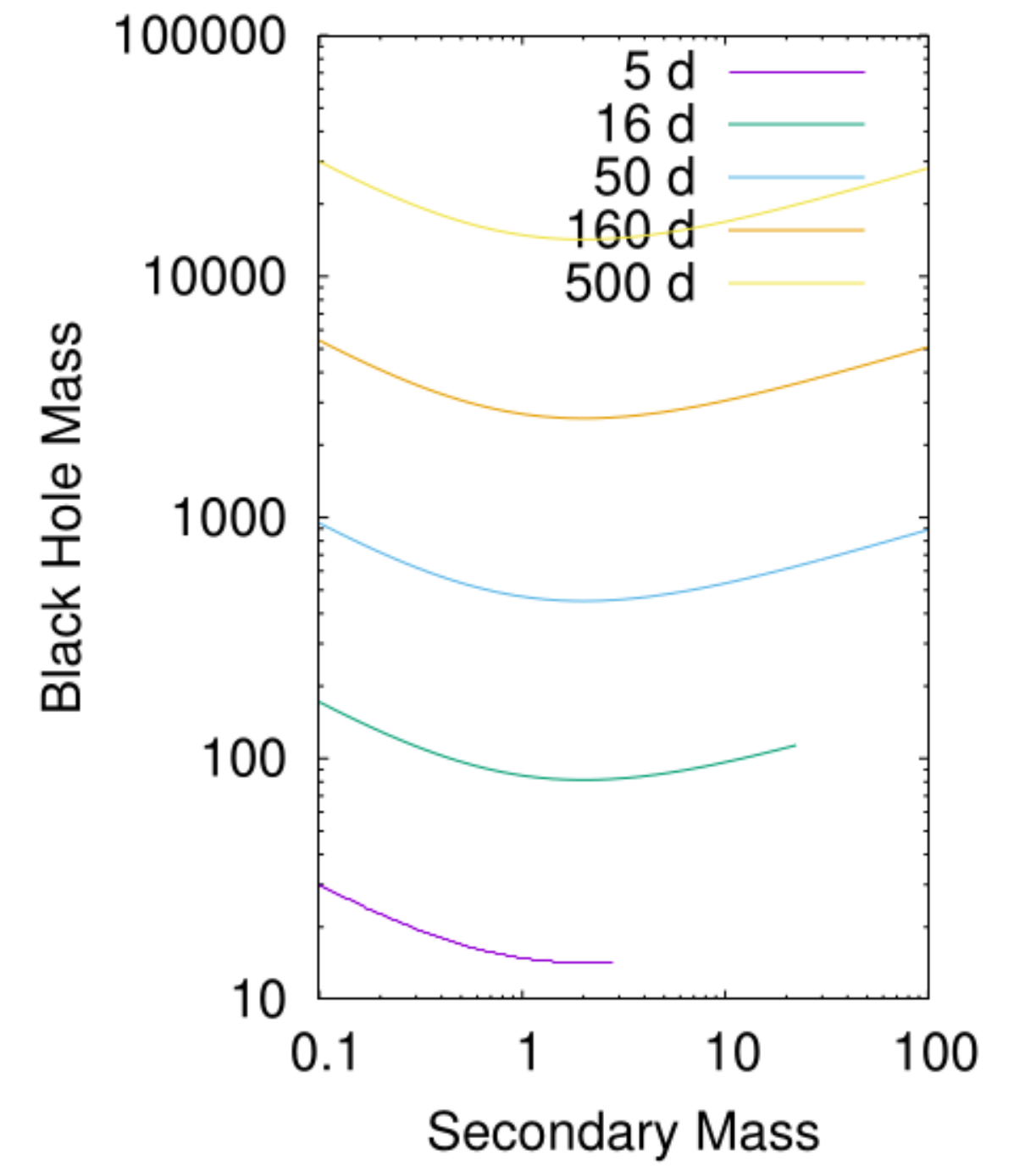}
	\caption{\label{mbhfig} Black hole mass as a function of the
	secondary mass if the secondary is on the main sequence with
	$\langle \rho \rangle$ given by Eq.~\ref{density}.  Curves are
	labeled by the precessional period.  All masses are in units of
	$M_\odot$.  Curves are only plotted for $q \le 0.2$ for which
	Eq.~\ref{periodfit} is valid.}
\end{figure}

If the secondary is on the zero-age main sequence its mean density may be
approximated
\begin{equation}
	\label{density}
	\langle \rho \rangle \approx {1 \over 2}
	\left({1 \over m}+{1 \over m^2}\right) \langle \rho \rangle_\odot,
\end{equation}
where $m \equiv m_s/M_\odot$ and $\langle \rho \rangle_\odot \approx 1.4$
g-cm$^{-3}$.  Combining these results, the black hole mass
$m_{BH} \equiv M_{BH}/M_\odot$ is
\begin{equation}
	\label{mBH}
	\begin{split}
		m_{BH} &\approx \left({G P_{pre}^2 \over
		6 \pi \times 6.86^2} {m+1 \over m^{2/3}}
		\langle \rho \rangle_\odot\right)^{3/4}\\
		&\approx 55 \left({P_{pre} \over \text{16.3 d}}\right)^{3/2}
		\left(m+1 \over m^{2/3}\right)^{3/4},
	\end{split}
\end{equation}
This is plotted in Fig.~\ref{mbhfig}.  Lower $M_{BH}$ are possible if the
secondary has evolved from the main sequence and its density has decreased
below the value of Eq.~\ref{density}.

\section{Discussion}
It is not evident how to test directly the hypothesis that FRB emerge from
the throat of an accretion disc.  In some cases \citep{K22b} indirect
arguments may be used, but these may be challenged as relying on unproved
models.  The results of this paper do not solve that problem because they
only imply relations between pairs of quantities ($m_{BH}$ and $m$, or
$m_{BH}$ and $q$, or $m_{BH}$ and $P_{orb}$, {\it etc.\/}), neither of which
is (at present) directly measurable, or between $P_{pre}$ and a quantity
($\langle \rho \rangle$, $P_{orb}$, $q$, $m$, {\it etc.\/}) that is not
directly measurable.  If many more periodically modulated FRB sources are
discovered it may be possible to compare their statistics to the predictions
of population synthesis and the model developed here.

A test of a leading alternative model, that FRB are produced by young
magnetars in young supernova remnants, may be possible, at least for one
source (FRB 121102) with periodically modulated activity \citep{R20}.
FRB 121102 has an extraordinarily large and rapidly decreasing rotation
measure \citep{M18,H21,P22,F23}.  In a na\"{\i}ve magnetar-SNR model, the
rotation measure decreases $\propto r^{-4} \propto t^{-4}$, where $r$ is the
radius of the SNR and $t$ is the time since the explosion, because the column
density and a frozen-in magnetic field each decrease $\propto r^{-2} \propto
t^{-2}$.  The observed rapid decrease would imply the source was formed
$\lesssim 10$ years before the first measurements of its rotation measure in
2016.  This power law decay would continue until only residual Galactic and
host galaxy contributions, orders of magnitude less than the observed
rotation measure, remain.

In the model proposed here, the rotation measure is produced by a chaotic
plasma environment of the accretion disc (and perhaps the disc itself), and
fluctuates irregularly, like the dispersion measure of FRB 190520B
\citep{K22c}.  In contrast to the power law decay predicted by the
magnetar-SNR model, the accretion disc model predicts that the rotation
measure may (but would not necessarily) reverse sign, or that its decay may
be replaced by a period of growth.  Continued observation of the rotation
measure of FRB 121102 thus may be a definitive test of the magnetar-SNR
model, even though it would not be a direct test of the accretion disc
model.
\section*{Data Availability}
This theoretical study did not obtain any new data.

\label{lastpage} 
\end{document}